\documentclass[a4paper,onecolumn]{article}
\usepackage[utf8]{inputenc}
\usepackage[T1]{fontenc}
\usepackage[USenglish]{babel}
\pdfoutput=1

\usepackage[a4paper]{geometry}
\usepackage[singlespacing]{setspace}

\usepackage{upgreek}

\usepackage[backend=biber,bibencoding=utf8,sorting=none,style=chem-acs,isbn=false,doi=false,url=false,eprint=false,articletitle=true]{biblatex}  
\AtEveryBibitem{\clearfield{note}}

\AtEveryBibitem{
 \clearlist{address}
 \clearlist{language}
 \clearfield{eprint}
 \clearfield{isbn}
 \clearfield{issn}
 \clearfield{series}
 \clearfield{note}
 \ifentrytype{journal article}{}{
    \clearlist{location}
 \clearfield{month}
 \clearfield{day}
 \clearfield{date}
 \clearfield{issue}
 \clearfield{doi}
  }
 }
 \DeclareSourcemap{%
  \maps[datatype=bibtex]{
    \map[overwrite]{
      \step[fieldsource=shortjournal]
      \step[fieldset=journaltitle,origfieldval]
    }
  }
}

\usepackage{xcolor}
\usepackage{graphicx}
\usepackage{tabularx}
\usepackage{multirow}
\usepackage{multicol}
\usepackage{amsmath}
\usepackage{amssymb}
\usepackage{amsfonts}
\usepackage{mathtools}

\usepackage{authblk}
\usepackage{xcolor}
\usepackage{comment}
\usepackage{cleveref}

\usepackage{booktabs,cancel,caption,colortbl,csquotes,helvet,mathpazo,listings,pgfplots}

\usepackage[range-phrase={\text{~to~}},output-decimal-marker={.}]{siunitx}
\DeclareSIUnit\photons{\textrm{ph}}

\usepackage[colorinlistoftodos,textsize=small]{todonotes}
\setuptodonotes{fancyline, color=green!40}
\usepackage[version=4]{mhchem}
\usepackage{lipsum}
\DeclareUnicodeCharacter{03B2}{\ensuremath{\beta}}

\date{}

\title{\textbf{
Coherent X-ray Scattering Reveals Nanoscale Fluctuations in Hydrated Proteins
}}

\begin{document}

\author[1]{Maddalena Bin}
\author[1]{Mario Reiser}
\author[1]{Mariia Filianina}
\author[1]{Sharon Berkowicz}
\author[1]{Sudipta Das}
\author[2]{Sonja Timmermann}
\author[3]{Wojciech Roseker}
\author[3,4]{Robert Bauer}
\author[1]{Jonatan \"{O}str\"{o}m}
\author[1]{Aigerim Karina}
\author[1,5,6]{Katrin Amann-Winkel}
\author[1]{Marjorie Ladd-Parada}
\author[3]{Fabian Westermeier}
\author[3]{Michael Sprung}
\author[7]{Johannes M\"{o}ller}
\author[3,8]{Felix Lehmkühler}
\author[2]{Christian Gutt}
\author[1]{Fivos Perakis\thanks{f.perakis@fysik.su.se}}

\affil[1]{Department of Physics, AlbaNova University Center, Stockholm University, 106 91 Stockholm, Sweden}
\affil[2]{Department Physik, Universit\"{a}t Siegen, Walter-Flex-Str. 3, 57072 Siegen, Germany}
\affil[3]{Deutsches Elektronen-Synchrotron, Notkestr. 85, 22607 Hamburg, Germany}
\affil[4]{Freiberg Water Research Center, Technische Universität Bergakademie Freiberg, 09599 Freiberg, Germany}
\affil[5]{Max-Planck-Institute for Polymer Research, 55128 Mainz, Germany}
\affil[6]{Institute of Physics, Johannes Gutenberg University, 55128 Mainz, Germany}
\affil[7]{European X-Ray Free-Electron Laser Facility, Holzkoppel 4, 22869 Schenefeld, Germany}
\affil[8]{The Hamburg Centre for Ultrafast Imaging, Luruper Chaussee 149, 22761 Hamburg, Germany}

\maketitle
\newpage

\section*{Abstract}
Hydrated proteins undergo a transition in the deeply supercooled regime, which is attributed to rapid changes in hydration water and protein structural dynamics. Here, we investigate the nanoscale stress relaxation in hydrated lysozyme proteins stimulated and probed by X-ray Photon Correlation Spectroscopy (XPCS). This approach allows us to access the nanoscale dynamic response in the deeply supercooled regime (${T = 180~}$K) which is typically not accessible through equilibrium methods. The relaxation time constants exhibit Arrhenius temperature dependence upon cooling with a minimum in the Kohlrausch-Williams-Watts exponent at ${T = 227~}$K. The observed minimum is attributed to an increase in dynamical heterogeneity, which coincides with enhanced fluctuations observed in the two-time correlation functions and a maximum in the dynamic susceptibility quantified by the normalised variance $\chi_T$. Our study provides new insights into X-ray stimulated stress relaxation and the underlying mechanisms behind spatio-temporal fluctuations in biological granular materials.

\begin{figure}[htbp]
    \includegraphics[width=0.5\textwidth]{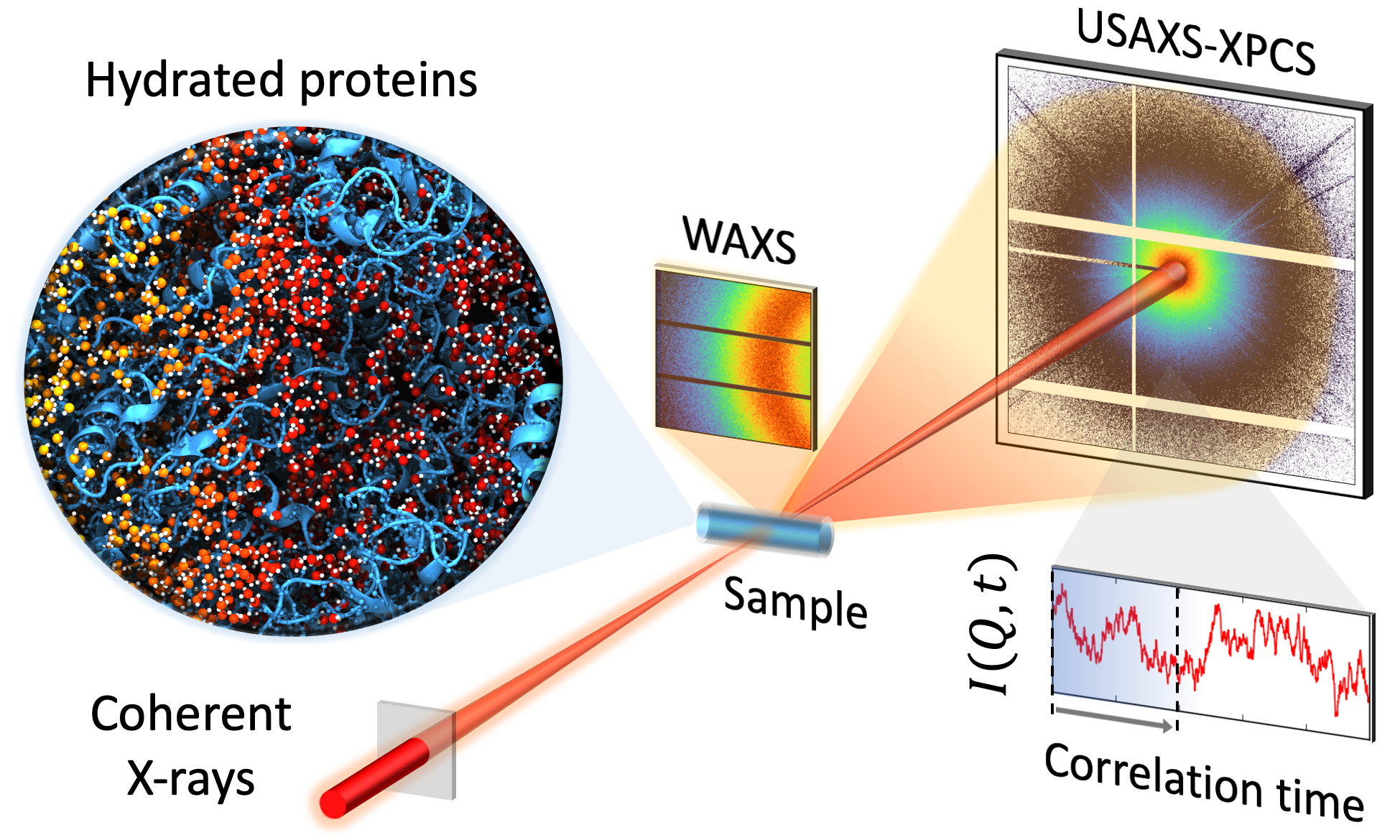}
    \centering
\end{figure}

\newpage

Proteins undergo a transition upon cooling below $T\approx230$~K which impacts their biological function~\cite{schiro_role_2019,ringe_the_2003}. A signature of a similar transition has also been observed for DNA~\cite{sokolov_slow_2001}, tRNA~\cite{roh_dynamics_2009,caliskan_dynamic_2006} and hydrated polymers~\cite{bailey_viscoelastic_2020,demichele_hysteresis_2019}. Despite the recent progress in the field, the origin of the transition is still a controversial topic and not fully understood. One hypothesis suggests that protein activity is reduced below the transition temperature due to the deactivation of certain degrees of freedom needed for structural transformations essential to protein function~\cite{doster_dynamical_1989, rasmussen_crystalline_1992, vitkup_solvent_2000,zaccai_how_2000,lee_microscopic_2001}. Nuclear magnetic resonance (NMR) studies indicate that motions relevant to protein functionality are activated above 230 K due to the “unfreezing” of hydration water~\cite{lewandowski_direct_2015}. This interpretation suggests that hydration water experiences an arrest of collective motion ($\alpha$-relaxation) upon cooling and that below this temperature, only the local motions are active ($\beta$-relaxation)~\cite{swenson_relaxation_2006}. As a result, this glass-like arrest of hydration water can lead to the deactivation of certain protein degrees of freedom relevant for biological activity. On the other hand, molecular dynamics simulations indicate that cold-denaturation in this temperature range can also lead to impairment of biological function due to protein unfolding~\cite{kim_computational_2016a,kozuch_low_2019,kim_computational_2016b,yang_a_2014}. In this case, the low-temperature denaturation occurs due to the disruption of the protein structure, which facilitates the intrusion of water into the protein’s interior and solvation of buried core hydrophobic residues. 

An alternative interpretation proposes that the hydration water is predominantly responsible for the observed low-temperature transition. It is postulated that liquid water exhibits a fragile-to-strong transition at $ T\approx230$~K as indicated by quasi-elastic neutron scattering (QENS) experiments on hydrated lysozyme powders~\cite{chen_observation_2006}. In this scenario, the low-temperature transition is triggered by changes in the hydration water dynamics, which in turn impact the protein activity. This hypothesis is linked to the proposed liquid-liquid transition in liquid water, which suggests a transition from a high-density to a low-density liquid (HDL and LDL)~\cite{kim_maxima_2017, kim_experimental_2020}. The liquid-liquid transition is hypothesized to take place in the deeply supercooled regime due to the existence of a liquid-liquid critical point~\cite{debenedetti_second_2020}. Simulations and experiments of proteins and other biomolecules in supercooled water indicate that the protein low-temperature transition can be associated with the liquid-liquid transition~\cite{kumar_glass_2006, mazza_more_2011, schiro_communication:_2013}. 

Experiments at highly coherent X-ray sources provide a unique opportunity to advance our understanding and gain new experimental insights into collective fluctuations during the low-temperature transition. X-ray photon correlation spectroscopy (XPCS) is a technique that utilizes coherent X-rays and can resolve collective nanoscale dynamics, ranging from microseconds to hours~\cite{madsen_structural_2016,grubel_x-ray_2008,lehmkuhler_from_2021}. XPCS has been demonstrated for a broad range of soft condensed matter systems~\cite{sandy_hard_2018, sinha_x-ray_2014}, including amorphous water where a liquid-liquid transition was observed in the ultraviscous regime~\cite{perakis_diffusive_2017}. However, due to experimental difficulties in working with radiation-sensitive samples, XPCS of protein systems became possible only recently with optimized experimental procedures~\cite{ragulskaya_interplay_2021,girelli_microscopic_2021,begam_kinetics_2021, reiser_resolving_2022, chushkin_probing_2022, vodnala_radiation_2016, vodnala_hard-sphere-like_2018,moron_gelation_2022,moller_x-ray_2019}.


Here, we explore the nanoscale dynamics in hydrated lysozyme powders from ambient to cryogenic conditions using XPCS. Hydrated protein powders allow to suppress freezing by confining water in the protein matrix. In this experiment we combine wide-angle X-ray scattering (WAXS) with XPCS in ultra-small-angle X-ray scattering geometry (USAXS). This approach provides insights into the previously unexplored low momentum transfer region associated with collective nanoscale fluctuations and stress relaxation, stimulated by the X-ray beam. The unique advantage of this approach is that it allows to resolve nanoscale dynamics deeply in the supercooled regime ($T=180$~K), which would normally be frozen in and therefore outside our experimental observation window.

\begin{figure}[htbp]
    \includegraphics[width=0.9\textwidth]{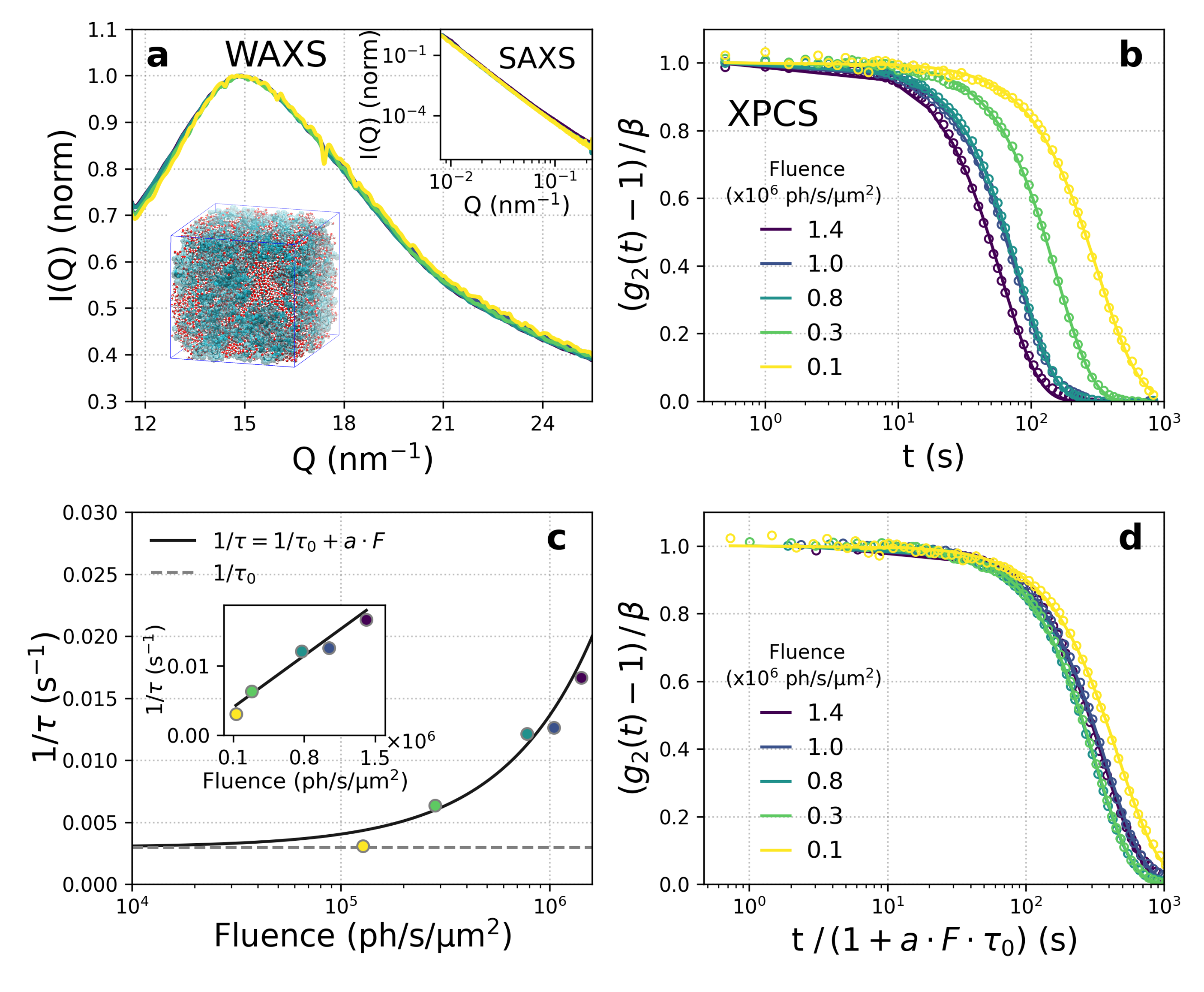}
    \centering
    \caption{\label{fig:fluence} (a) Wide- and small-angle X-ray scattering (WAXS and inset SAXS) intensity as a function of momentum transfer $Q$ for different fluences at room temperature ($T=300$~K). The schematic depicts the hydrated protein matrix~\cite{bin_wide-angle_2021}. (b) Intensity autocorrelation functions $g_2$ at momentum transfer $Q=0.1~{\rm nm^{-1}}$ for different fluences (in units of photons/second/area), as indicated in the legend. (c) The decay rate 1/$\tau$ extracted from the fit of the $g_2$ functions (solid lines) for variable fluence. The inset shows the same data on linear scale. (d) The renormalised intensity autocorrelation functions $g_2$ at momentum transfer $Q=0.1~{\rm {\rm nm^{-1}}}$. The time axis is normalised to the corresponding fluence $F$ by calculating t${/ (1+a\cdot F \cdot } \tau_0{\rm )}$, where $\tau_0$ is the equilibrium time constant extracted by extrapolation to $F$ = 0.
    }
\end{figure}

In order to unravel the origin for the observed dynamics we perform fluence-dependent measurements at room temperature ($T=300$~K). \Cref{fig:fluence}a shows the WAXS and SAXS intensity as a function of momentum transfer for different fluences. A minor shift of the momentum transfer Q is observed in the WAXS region, whereas no significant changes in the SAXS. In addition, for a given fluence the WAXS signal does not exhibit any significant changes as a function of measurement time (see SI). \Cref{fig:fluence}b shows the temporal intensity autocorrelation functions $g_2$ at different fluences indicated in the legend, corresponding from 0\% (blue) to 90\% attenuation (yellow). The $g_2$ function is defined as~\cite{berne_dynamic_2000}

\begin{equation}
    g_2(Q,t) = \frac{\langle I(Q,t_0) I(Q,t_0+t) \rangle}{\langle I(Q,t_0) \rangle^2} \,.
    \label{g2}
\end{equation}

Here, $I(Q,t_0)$ and $I(Q,t_0+t)$ denote the intensity of a pixel at time $t_0$ and after delay time $t$, respectively. The bracket notation refers to averaging over time $t_0$ and pixels that belong to a given momentum transfer $Q$-bin, i.e. a thin annulus slice around the beam center corresponding to similar momentum transfers $Q$. The momentum transfer $Q$ is defined as $Q=4\pi/\lambda\,\sin(2\theta)$, where $\lambda$ is the wavelength and $2\theta$ the scattering angle. A stretched exponential function (solid line) is fitted to the resulting correlation functions

\begin{equation}
    g_2 (Q,t) = \beta(Q)\,\exp\bigl[-2(t/\tau(Q))^{\alpha(Q)} \bigr] + c\,,
    \label{g2-fit}
\end{equation}

where $\beta$ is the speckle contrast, $c$ is the baseline, $\tau$ is the time constant and $\alpha$ is the Kohlrausch-Williams-Watts (KWW) exponent~\cite{williams_non-symmetrical_1970}. The obtained $g_2$ functions exhibit an acceleration of the dynamics for higher fluences (color-coded in the legend). 

Based on the fluence dependence, the observed dynamics are attributed to nanoscale stress relaxation stimulated by the X-ray beam, which is related to the intrinsic dynamic viscoelastic response of the system to external stimulus. This approach of probing the system response function to external stimuli resembles conceptually the use of cyclic shearing, which can provide information about the dynamic behavior of granular media close to the "jamming transition" \cite{dauchot_dynamical_2005}, as well as dielectric spectroscopy which can shed light into the dynamic system response stimulated by external fields near the glass transition~\cite{albert_fifth-order_2016}. The time constants extracted from the intensity autocorrelation functions $g_2$ are inversely proportional to the fluence and resemble ballistic motion (see SI). Similar observations in oxide glasses indicate that the beam-driven dynamics are due to radiolysis-induced atomic displacement~\cite{pintori_relaxation_2019,ruta_hard_2017}, whereas numerical simulations of colloidal particles indicate that the origin of the ballistic motion can be linked to elastic relaxation~\cite{bouzid_elastically_2017}.
The time constant $\tau$ extracted from the fits as a function of fluence is shown in \Cref{fig:fluence}c. The solid line depicts the relation between the extrapolated equilibrium time constant $\tau_0=336\pm110$ s and the measured relaxation time $\tau$ which is modelled by $ 1/\tau = 1/\tau_0 +a\cdot F$. The constant $a$ couples the system's dynamic response to the X-ray beam, which here is estimated from the fit as $a = (1.1 \pm 0.2)\times10^{-8} $ $\upmu$m$^2$/ph, while $F$ is the fluence in ph/s/$\upmu$m$^2$.
Another similarity to oxide glasses is related to the KWW exponent, which here is  $\alpha \approx 1.5$ and similar to those obtained for oxide glasses~\cite{pintori_relaxation_2019,ruta_hard_2017}. However, here the probed Q-range is not directly sensitive to the local molecular rearrangements but reflects instead the stimulated dynamic response over nanometer length scales. The scattering intensity arises from the density difference in the nanoscale protein grain boundaries and the observed dynamics can be attributed to collective stress relaxation, as the system converts from a jammed granular state to elastically driven regime~\cite{jose_similarities_2012, bouzid_elastically_2017}.

In hydrated protein-based systems, this kind of dynamic behavior can be potentially influenced by radiation damage attributed to the reaction of proteins with the radicals produced by radiolysis, such as OH radicals~\cite{schwarz_applications_1969}. Such effects however depend on protein concentration as a highly solvent accessible environment is more susceptible to OH radicals~\cite{kuwamoto_radiation_2004} and therefore are more significant in the dilute regime than in hydrated powders. Furthermore, in the present experiment the estimated temperature rise is below 1~K (see SI) which is consistent with the observed shift in the WAXS~\cite{bin_wide-angle_2021} (see \Cref{fig:fluence}a), and is insufficient for inducing any changes due to thermal denaturation. 
The observed dynamics are isotropic (see SI), except from the presence of streaks due to the grain boundaries that had to be masked as previously~\cite{perakis_diffusive_2017}.

By normalising the correlation function $g_2$ time axis with respect to the fluence shown in \Cref{fig:fluence}d, we observe that the curves overlap independent of the fluence used, as seen previously for oxide glasses~\cite{ruta_hard_2017}. In the current system however, this extrapolation is limited to room-temperature data; that is because the dynamics in the low momentum transfer range probed here would be too slow and outside the experimental window for the deeply supercooled regime. By stimulating stress relaxation with the X-rays we are able to obtain information about the nanoscale dynamic response even at cryogenic temperatures.

\begin{figure}[htbp]
    \includegraphics[width=1\textwidth]{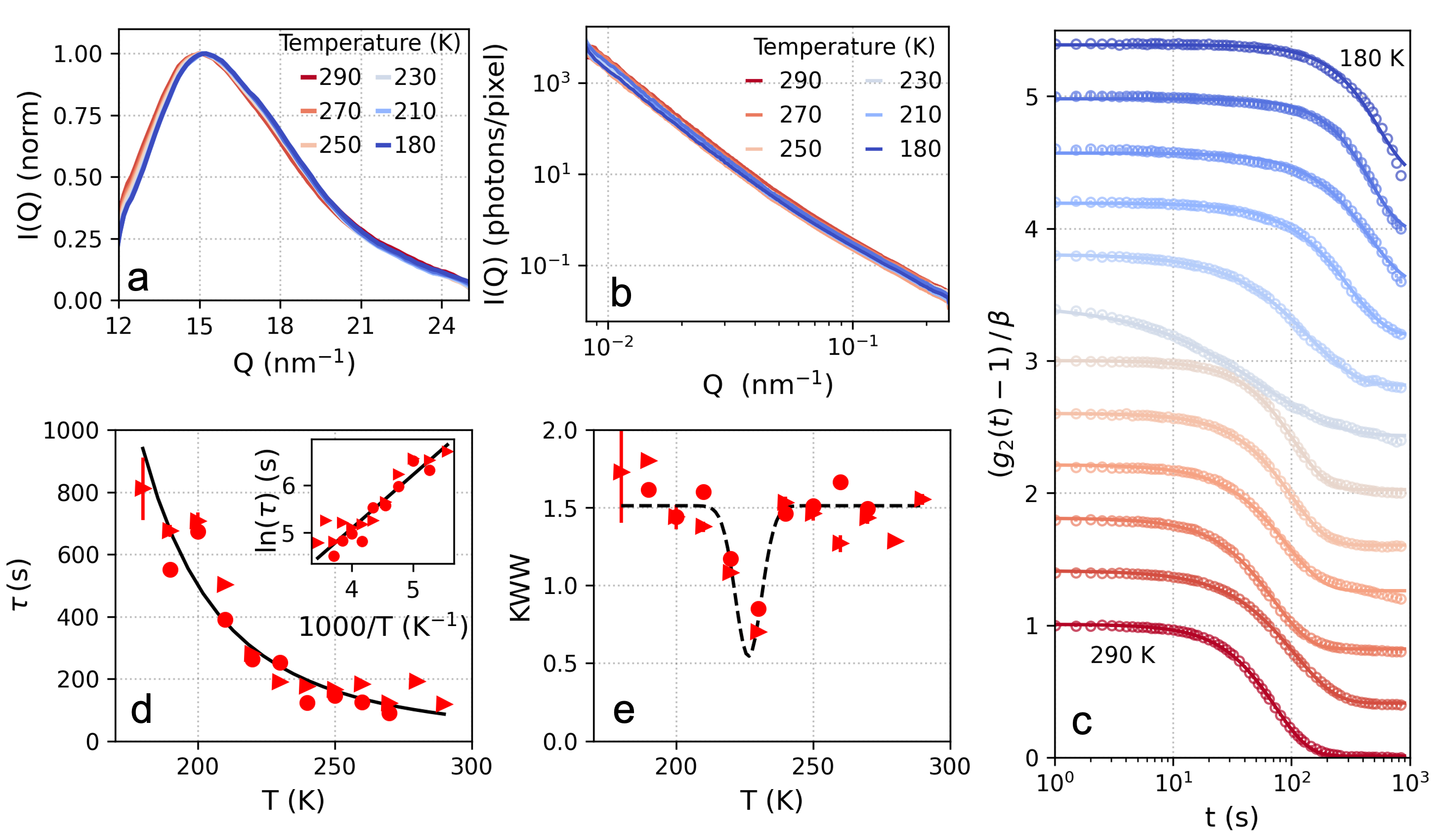}
    \caption{\label{fig:temp} Temperature-dependent measurements. (a) The WAXS and (b) SAXS scattering intensity at different temperatures indicated in the legend. (c) Intensity autocorrelation functions $g_2$ for different temperatures, upon cooling from $T=290$~K to $T=180$~K, as indicated in the plot. The data shown are calculated at momentum transfer $Q=0.1~{\rm nm^{-1}}$ and the solid lines indicate the fits with a stretched exponential. An offset has been added to facilitate the comparison. (d) The time constants $\tau$ extracted from the fits in panel c. The inset shows the  logarithm of the time constant $\tau$ as a function of the inverse temperature $10^3/T$ where the solid line indicates the Arrhenius fit. (e) The Kohlrausch-Williams-Watts (KWW) exponent as a function of temperature. The dashed line is a guide to the eye (Gaussian fit) indicating a minimum at $T=227$~K. The various symbols in panels (d) and (e) indicate data acquired during different beamtimes with similar conditions (see methods).}
\end{figure}

By recording simultaneously X-ray diffraction data in WAXS geometry along with the XPCS measurements we ensure that the samples have not crystallised (\Cref{fig:temp}a, see also SI). The observed changes in WAXS intensity as a function of temperature, such as the shift towards larger Q upon cooling, agree with previous investigations and indicate temperature-dependent changes on atomic lengthscales \cite{bin_wide-angle_2021}. We do not observe significant temperature-dependent changes in the SAXS region, as shown in the \Cref{fig:temp}b.
The temperature dependence of the dynamics was measured by using $F =1.5\cdot10^6$~ph/s/$\upmu$m$^2$ upon cooling from $T=290$~K down to $T=180$~K. This experimental condition corresponds to a dose rate of $1.58$~kGy/s (see SI). The intensity autocorrelation $g_2$ functions are visualized in \Cref{fig:temp}. The $g_2$ functions indicate that the system exhibits a slowing down of the dynamics upon cooling. 

The extracted time constants $\tau$ are shown as a function of temperature in \Cref{fig:temp}d, where the solid line depicts an Arrhenius fit using the relation $\tau(T) = A\cdot e^{E_a/k_BT}$, where $A=1.8\pm1.1$~s is the amplitude and $E_A = 9.4 \pm 1.1$ kJ/mol is the activation energy. Additionally, the natural logarithm of the time constants $\tau$ are depicted in \Cref{fig:temp}d as a function of the inverse of the temperature (Arrhenius plot, inset). The Arrhenius analysis yields an activation energy which is comparable with that obtained by QENS (13 kJ/mol)~\cite{chen_observation_2006}. The difference here is that the QENS measurements were performed at a higher momentum transfer region, reflecting local molecular diffusion, whereas the low momentum transfer $Q$ probed here reflects nanoscale viscoelastic motion. This difference could explain why here any noticeable crossover in the proximity of the low-temperature transition temperature is not observed, indicated by QENS in hydrated lysozyme powders~\cite{chen_observation_2006}. This result is consistent with dielectric spectroscopy measurements, which showed no sign of a transition in the temperature dependence of conductivity~\cite{pawlus_conductivity_2008} in hydrated lysozyme proteins. 

The corresponding KWW exponents as a function of temperature are shown in \Cref{fig:temp}e. We observe that the KWW exponent exhibits a minimum from an average value of $\alpha \approx 1.5$ to $\alpha < 1$ at $ T = 227$~K. Similar transitions have been attributed to the emergence of dynamical heterogeneities upon approaching the glass transition temperature~\cite{ruta_wave-vector_2020,jain_three-step_2022,frenzel_glass-liquid_2021}, although here we observe that the exponent values return to $\alpha \approx 1.5$ below $T=210$~K. 
This behaviour can also be seen directly from the lineshape of the $g_2$ functions, which appear distinctly more stretched at $T\approx230$ K. The observed minimum was reproduced over several beamtimes with similar experimental conditions, indicated by the different symbols in \Cref{fig:temp}d and \Cref{fig:temp}e (see Experimental Methods section).

Calculating the two-time correlation (TTC) function~\cite{madsen_beyond_2010} allows us to quantify the dynamical heterogeneity. The TTC is defined as:
\begin{equation}
    c_2(Q,t_1,t_2) = 
    \frac{
        \langle I(Q,t_1) I(Q,t_2) \rangle_{\text{pix}}}
        {\langle I(Q,t_1)\rangle_{\text{pix}} \langle I(Q,t_2)\rangle_{\text{pix}}}\,,
    \label{c2}
\end{equation}

where $I(Q,t_1)$ and $I(Q,t_2)$ denote the intensity of a pixel at distinct times $t_1$ and $t_2$. The subscript `pix' implies that, contrary to the $g_2$ definition, the averaging is, in this case, solely performed over pixels within the same Q-bin and not over time.

\begin{figure}[htbp]
    \includegraphics[width=0.8\textwidth]{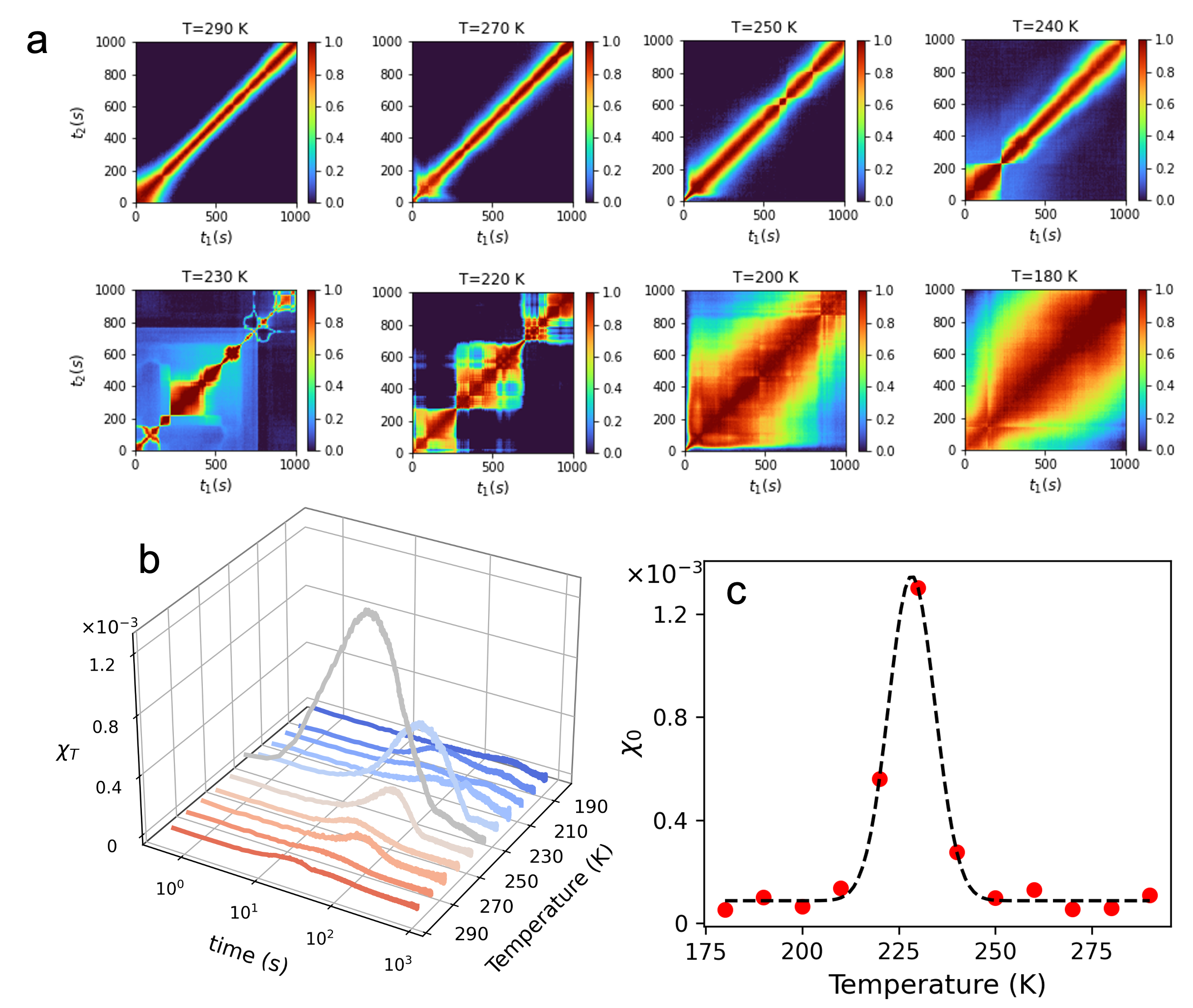}
    \centering
    \caption{\label{fig:chi4} 
    (a) Two-time correlation function (TTC) at different temperatures indicated in the panel titles at momentum transfer $Q=0.1~{\rm nm^{-1}}$. (b) The normalised variance $\chi_T$ at different temperatures extracted from the TTC. (c) The maximum of the normalised variance $\chi_0$ obtained at different temperatures indicates a maximum at $T=227$~K.}
\end{figure}

In~\Cref{fig:chi4}a the TTC functions for temperatures ranging from $T=290$~K to $T=180$~K are shown. We observe that overall the TTC lineshape for higher temperatures looks smooth and continuous. In some instances, an initial acceleration (see e.g. $T=270$~K or $T=250$~K) or deceleration (see e.g. $T=290$~K) is observed before the lineshape is stabilised, which can be attributed to the initial interaction and stress relaxation stimulated by the beam. In addition, pronounced fluctuations manifest for $T=230$~K. These fluctuations are enhanced for the range mainly between $T=230$ - $220$~K and diminish below $T=210$~K. It is worth noting that the KWW exponent minimum observed in~\Cref{fig:temp} is a direct consequence of averaging the TTC over multiple correlation times.

The dynamical heterogeneity is quantified by calculating the normalized variance $\chi_T$ of the TTC function, which is an experimentally accessible estimator of the four-point dynamical susceptibility~\cite{berthier_direct_2005}. The variance is calculated from the TTC, by using the relation

\begin{equation}
    \chi_T(Q, \delta t)=\frac{\langle c_2(Q, t, \delta t)^2 \rangle_{ t}-\langle c_2(Q, t, \delta t) \rangle_{ t}^2}{\langle c_2(Q, t, \delta t=0) \rangle_{ t}^2}
\end{equation}

where $t$ and $\delta t$ correspond to the diagonal and anti-diagonal axes of the TTC. The subscript $t$ implies that the averaging is, in this case, performed over the TTC diagonal axis, as previously~\cite{perakis_diffusive_2017}. 

The $\chi_T$ calculated at different temperatures is shown in \Cref{fig:chi4}b. We observe that the amplitude of $\chi_T$ is maximised at $T=230$~K, consistent with the enhancement of the observed fluctuations in the TTC, which result in distinct peaks in the $\chi_T$ distribution. Interestingly, the mean relaxation time at the $\chi_T$ peak is significantly shorter than the average relaxation time at $T=230$~K (26 s and 220 s respectively), indicating that the observed fluctuations can be attributed to a faster intrinsic dynamic process. The maximum value of the $\chi_T$, denoted as $\chi_0$ in \Cref{fig:chi4}c, exhibits a maximum at $T=227$~K. This enhancement in $\chi_T$ is an indication of maximization of dynamical heterogeneities in this temperature range which manifests as fluctuations in the TTC function. Such enhancement of the dynamic susceptibility in granular materials has been previously associated with a growing dynamic correlation length due to spatio-temporal fluctuations~\cite{dauchot_dynamical_2005,keys_measurement_2007}.  

The present observations are consistent with previous investigations of the dynamics in hydrated lysozyme, where a maximum in the specific heat capacity was attributed to a sharp change of local order and enhanced hydrogen bond fluctuations~\cite{mazza_more_2011, mallamace_nmr_2008}. Furthermore, a maximum in the dynamic susceptibility at $ T = 230$~K has been associated with growth in size of dynamic heterogeneity in confined water~\cite{zhang_dynamic_2009}. The current data are also in-line with the crossing of the Widom line, defined as the locus of points in the P-T surface, which have a maximum in the correlation length. The specific heat capacity and isothermal compressibility of pure supercooled water exhibit maxima at the Widom line~\cite{kim_maxima_2017, pathak_enhancement_2021}, which are thermodynamic response functions representing enthalpy and density fluctuations. Even though the confinement of hydration water in the protein matrix can influence the local structure~\cite{bin_wide-angle_2021} our data here indicate that it is still possible to capture collective fluctuations at $ T = 227$~K which correlates with the Widom line temperature at ambient pressure in pure liquid water~\cite{kim_maxima_2017, pathak_enhancement_2021}.

Summarising, we have studied the stress relaxation dynamics stimulated by X-rays in hydrated lysozyme powder by using USAXS-XPCS. This approach allows us to obtain information about the nanoscale stimulated dynamic response down to deeply supercooled conditions ($T=180$~K) which is inaccessible under equilibrium conditions. The extracted time constants exhibit Arrhenius temperature dependence accompanied by a sharp minimum in the KWW exponent  $ T = 227$~K, indicative of dynamic heterogeneity. TTC analysis indicates the presence of pronounced dynamic fluctuations, which are maximised at the same temperature range. The results are consistent previously observed maxima in the specific heat capacity in hydrated lysozyme powder~\cite{mazza_more_2011, mallamace_nmr_2008} attributed to enhanced density and enthalpy fluctuations. From a general point of view, our study paves the way for future experiments following nanoscale fluctuations and stress relaxation in systems where equilibrium dynamics are not accessible with standard methods, such as for granular matter and glassy materials.

\section*{Experimental Methods}

The lysozyme protein used is lyophilized powder from chicken egg white purchased from Sigma-Aldrich (L6876).
The powder was grinded with a mortar to reduce the grain size, and used without further purification or drying process.
Lysozyme powder was hydrated by exposing it to water vapor in a closed hydration chamber, controlling the humidity and the exposure time to reach the desired hydration value, as characterised previously~\cite{bin_wide-angle_2021}.
At hydration level below $h\approx0.3$ crystallization is suppressed due to the confinement of water in the protein matrix~\cite{rupleyj.a_protein_1991}. The hydration level for each sample was characterised during the hydration process by measuring the weight before and after hydration. The data shown correspond to an average hydration level of $h=0.28\pm0.05$.

The data were acquired at the Coherence Applications beamline P10 at PETRA III (proposal numbers I-20200072 EC and I-20220280 EC) at the Deutsches Elektronen-Synchrotron (DESY). The measurements were performed simultaneously in ultra-small-angle X-ray scattering ($\text{USAXS}$) and wide-angle X-ray scattering (WAXS) geometries  using a Si(111) monochromator. The experimental results were repeated and reproduced under similar condition; for details on the experimental parameters refer to~\Cref{tab:exp-param}. For the two beamtimes we used different sample environments including a Linkam Stage (model HFSX350) and a liquid-nitrogen cold-finger cryostat in vacuum. For measuring USAXS the Eiger detector was located at a distance of 21.2 m from the sample, while the Pilatus 300k detector was used to capture the WAXS signal. The hydrated protein samples were filled into quartz capillaries of 1.5 mm in diameter and the temperature was controlled in order to have a cooling rate of 5~K/min. By tracking the scattering intensity in WAXS it was possible to monitor freezing of the sample (see SI). In addition, preliminary XPCS measurements were carried out at beamline ID02 at the European Synchrotron Radiation Facility (ESRF). 

\begin{table}[tbhp]
  \centering
  \caption{The experimental X-ray parameters used for the two experiments, including the proposal number, photon energy, beam size, flux, sample environment, detector and sample-detector distance (SDD) for SAXS and WAXS geometries. }\label{tab:exp-param}
  \begin{tabular}{lcc}
    \hline
    Experiment    & I-20200072 EC   &  I-20220280 EC    \\ 
    \hline
    Energy (keV)  & 12.4  & 9.0   \\
    Beam size ($\upmu$m$^2$) & $30\times30$   & $30\times30$  \\
    Flux (x$10^9$\,ph/s)& 4.0 & 6.0   \\
    Sample environment  & Linkam stage  & Cryostat  \\
    SAXS detector  &    Eiger 500k  & Eiger 4M  \\
    SAXS SDD (m) &  21.2    & 21.2  \\
    WAXS detector   &    Pilatus 300k & Pilatus 300k\\
    WAXS SDD (m) & 0.21 &0.20   \\
    \hline
  \end{tabular}
\end{table}

\newpage
\section*{Acknowledgements}
We acknowledge financial support by the Swedish National Research Council (Vetenskapsr\text{\aa}det) under Grant No. 2019-05542 and within the R\"{o}ntgen-\text{\AA}ngstr\"{o}m Cluster Grant No. 2019-06075. This research is supported by Center of Molecular Water Science (CMWS) of DESY in an Early Science Project, the MaxWater initiative of the Max-Planck-Gesellschaft (project No. CTS21:1589), Carl Tryggers and the Wenner-Gren Foundations (project No. UPD2021-0144). Parts of this research were carried out at the light source PETRA III at DESY, a member of the Helmholtz Association (HGF). FL is supported by the Cluster of Excellence ‘Advanced Imaging of Matter’ of the Deutsche Forschungsgemeinschaft (DFG)–EXC 2056–project ID 390715994. KAW acknowledges Ragnar Söderberg Stiftelse for finacial support. CG and ST acknowledge funding by BMBF 05K19PS1, 05K20PSA and 05K22PS1. We acknowledge the European Synchrotron Radiation Facility (ESRF) for provision of synchrotron radiation facilities, and we thank Theyencheri Narayanan and Thomas Zinn for assistance with preliminary measurements at the beamline ID02.

\section*{Supporting information}
See SI for more details on the detection of crystallization, momentum transfer and azimuthal dependence of the time constants,  as well as on the temperature increase and dose estimations, including additional references~\cite{burtonlhenke_x-ray_1993,yang_protein-water_1979,leung_an_1999,fujiwara_measurements_2017,lehmkuhler_emergence_2020a}.   

\section*{Note}
The authors declare no competing financial interests.

\section*{Data availability}
The data that support the findings of this study are available from the corresponding author upon request. 

\printbibliography

@article{bin_wide-angle_2021,
	title = {Wide-angle {X}-ray scattering and molecular dynamics simulations of supercooled protein hydration water},
	volume = {23},
	copyright = {All rights reserved},
	issn = {1463-9084},
	url = {https://pubs.rsc.org/en/content/articlelanding/2021/cp/d1cp02126e},
	doi = {10.1039/D1CP02126E},
	language = {en},
	number = {34},
	urldate = {2021-09-03},
	journal = {Phys. Chem. Chem. Phys.},
	author = {Bin, Maddalena and Yousif, Rafat and Berkowicz, Sharon and Das, Sudipta and Schlesinger, Daniel and Perakis, Fivos},
	month = sep,
	year = {2021},
	pages = {18308--18313},
}

@article{schiro_role_2019,
	title = {Role of hydration water in the onset of protein structural dynamics},
	volume = {31},
	issn = {0953-8984},
	url = {https://doi.org/10.1088/1361-648x/ab388a},
	doi = {10.1088/1361-648X/ab388a},
	language = {en},
	number = {46},
	urldate = {2021-08-26},
	journal = {J. Phys.: Condens. Matter},
	author = {Schirò, Giorgio and Weik, Martin},
	month = aug,
	year = {2019},
	pages = {463002},
}

@article{jose_similarities_2012,
	title = {Similarities between protein folding and granular jamming},
	volume = {3},
	copyright = {2012 The Author(s)},
	issn = {2041-1723},
	url = {https://www.nature.com/articles/ncomms2177},
	doi = {10.1038/ncomms2177},
	language = {en},
	number = {1},
	urldate = {2021-03-17},
	journal = {Nat. Commun.},
	author = {Jose, Prasanth P. and Andricioaei, Ioan},
	month = oct,
	year = {2012},
	pages = {1161},
}

@article{kozuch_low_2019,
	title = {Low temperature protein refolding suggested by molecular simulation},
	volume = {151},
	issn = {0021-9606},
	url = {https://aip-scitation-org.ezp.sub.su.se/doi/full/10.1063/1.5128211},
	doi = {10.1063/1.5128211},
	number = {18},
	urldate = {2021-03-02},
	journal = {J. Chem. Phys.},
	author = {Kozuch, Daniel J. and Stillinger, Frank H. and Debenedetti, Pablo G.},
	month = nov,
	year = {2019},
	pages = {185101},
}

@article{kim_computational_2016b,
	title = {Computational investigation of cold denaturation in the {Trp}-cage miniprotein},
	volume = {113},
	issn = {0027-8424, 1091-6490},
	url = {https://www.pnas.org/content/113/32/8991},
	doi = {10.1073/pnas.1607500113},
	language = {en},
	number = {32},
	urldate = {2021-03-02},
	journal = {Proc. Natl. Acad. Sci. U.S.A},
	author = {Kim, Sang Beom and Palmer, Jeremy C. and Debenedetti, Pablo G.},
	month = aug,
	year = {2016},
	pages = {8991--8996},
}

@article{ringe_the_2003,
	series = {Walter {Kauzmann}`s 85th {Birthday}},
	title = {The ‘glass transition’ in protein dynamics: what it is, why it occurs, and how to exploit it},
	volume = {105},
	issn = {0301-4622},
	shorttitle = {The ‘glass transition’ in protein dynamics},
	url = {https://www.sciencedirect.com/science/article/pii/S0301462203000966},
	doi = {10.1016/S0301-4622(03)00096-6},
	language = {en},
	number = {2},
	urldate = {2021-02-26},
	journal = {Biophys. Chem.},
	author = {Ringe, Dagmar and Petsko, Gregory A.},
	month = sep,
	year = {2003},
	pages = {667--680},
}

@article{demichele_hysteresis_2019,
	title = {Hysteresis in the temperature dependence of the {IR} bending vibration of deeply cooled confined water},
	volume = {150},
	issn = {0021-9606},
	url = {https://aip.scitation.org/doi/full/10.1063/1.5096988},
	doi = {10.1063/1.5096988},
	number = {22},
	urldate = {2021-02-11},
	journal = {J. Chem. Phys.},
	author = {De Michele, Vincenzo and Levantino, Matteo and Cupane, Antonio},
	month = jun,
	year = {2019},
	pages = {224509},
}

@article{pathak_enhancement_2021,
	title = {Enhancement and maximum in the isobaric specific-heat capacity measurements of deeply supercooled water using ultrafast calorimetry},
	volume = {118},
	issn = {0027-8424, 1091-6490},
	url = {https://www.pnas.org/content/118/6/e2018379118},
	doi = {10.1073/pnas.2018379118},
	language = {en},
	number = {6},
	urldate = {2021-02-02},
	journal = {Proc. Natl. Acad. Sci. U.S.A},
	author = {Pathak, Harshad and Späh, Alexander and Esmaeildoost, Niloofar and Sellberg, Jonas A. and Kim, Kyung Hwan and Perakis, Fivos and Amann-Winkel, Katrin and Ladd-Parada, Marjorie and Koliyadu, Jayanath and Lane, Thomas J. and Yang, Cheolhee and Lemke, Henrik Till and Oggenfuss, Alexander Roland and Johnson, Philip J. M. and Deng, Yunpei and Zerdane, Serhane and Mankowsky, Roman and Beaud, Paul and Nilsson, Anders},
	month = feb,
	year = {2021},
}

@article{lehmkuhler_emergence_2020a,
	title = {Emergence of anomalous dynamics in soft matter probed at the {European} {XFEL}},
	volume = {117},
	copyright = {© 2020 . https://www.pnas.org/site/aboutpnas/licenses.xhtmlPublished under the PNAS license.},
	issn = {0027-8424, 1091-6490},
	url = {https://www.pnas.org/content/117/39/24110},
	doi = {10.1073/pnas.2003337117},
	language = {en},
	number = {39},
	urldate = {2021-01-25},
	journal = {Proc. Natl. Acad. Sci. U.S.A},
	author = {Lehmkühler, Felix and Dallari, Francesco and Jain, Avni and Sikorski, Marcin and Möller, Johannes and Frenzel, Lara and Lokteva, Irina and Mills, Grant and Walther, Michael and Sinn, Harald and Schulz, Florian and Dartsch, Michael and Markmann, Verena and Bean, Richard and Kim, Yoonhee and Vagovic, Patrik and Madsen, Anders and Mancuso, Adrian P. and Grübel, Gerhard},
	month = sep,
	year = {2020},
	pmid = {32934145},
	note = {Publisher: National Academy of Sciences
Section: Physical Sciences},
	pages = {24110--24116},
}

@article{moller_x-ray_2019,
	title = {X-ray photon correlation spectroscopy of protein dynamics at nearly diffraction-limited storage rings},
	volume = {6},
	copyright = {https://creativecommons.org/licenses/by/4.0/},
	issn = {2052-2525},
	url = {https://journals.iucr.org/m/issues/2019/05/00/tj5024/},
	doi = {10.1107/S2052252519008273},
	language = {en},
	number = {5},
	urldate = {2021-01-11},
	journal = {IUCrJ},
	author = {Möller, J. and Sprung, M. and Madsen, A. and Gutt, C.},
	month = sep,
	year = {2019},
	note = {Number: 5
Publisher: International Union of Crystallography},
	pages = {794--803},
}

@article{swenson_relaxation_2006,
	title = {Relaxation {Processes} in {Supercooled} {Confined} {Water} and {Implications} for {Protein} {Dynamics}},
	volume = {96},
	url = {https://link.aps.org/doi/10.1103/PhysRevLett.96.247802},
	doi = {10.1103/PhysRevLett.96.247802},
	number = {24},
	urldate = {2020-09-04},
	journal = {Phys. Rev. Lett.},
	author = {Swenson, Jan and Jansson, Helén and Bergman, Rikard},
	month = jun,
	year = {2006},
	pages = {247802},
}

@article{rupleyj.a_protein_1991,
	title = {Protein hydration and function},
	volume = {41},
	language = {en},
	journal = {Adv. Protein Chem.},
	author = {{Rupley, J.A} and {Careri, G.}},
	year = {1991},
	pages = {37--172},
}

@article{perakis_diffusive_2017,
	title = {Diffusive dynamics during the high-to-low density transition in amorphous ice},
	volume = {114},
	issn = {0027-8424, 1091-6490},
	url = {http://www.pnas.org/lookup/doi/10.1073/pnas.1705303114},
	doi = {10.1073/pnas.1705303114},
	language = {en},
	number = {31},
	urldate = {2020-07-09},
	journal = {Proc. Natl. Acad. Sci. U.S.A.},
	author = {Perakis, Fivos and Amann-Winkel, Katrin and Lehmkühler, Felix and Sprung, Michael and Mariedahl, Daniel and Sellberg, Jonas A. and Pathak, Harshad and Späh, Alexander and Cavalca, Filippo and Schlesinger, Daniel and Ricci, Alessandro and Jain, Avni and Massani, Bernhard and Aubree, Flora and Benmore, Chris J. and Loerting, Thomas and Grübel, Gerhard and Pettersson, Lars G. M. and Nilsson, Anders},
	month = aug,
	year = {2017},
	pages = {8193--8198},
}

@article{kim_computational_2016a,
	title = {Computational investigation of dynamical transitions in {Trp}-cage miniprotein powders},
	volume = {6},
	issn = {2045-2322},
	url = {https://www.nature.com/articles/srep25612},
	doi = {10.1038/srep25612},
	language = {en},
	number = {1},
	urldate = {2020-02-20},
	journal = {Sci. Rep.},
	author = {Kim, Sang Beom and Gupta, Devansh R. and Debenedetti, Pablo G.},
	month = may,
	year = {2016},
	pages = {1--8},
}

@article{zaccai_how_2000,
	title = {How {Soft} {Is} a {Protein}? {A} {Protein} {Dynamics} {Force} {Constant} {Measured} by {Neutron} {Scattering}},
	volume = {288},
	issn = {0036-8075, 1095-9203},
	shorttitle = {How {Soft} {Is} a {Protein}?},
	url = {https://science.sciencemag.org/content/288/5471/1604},
	doi = {10.1126/science.288.5471.1604},
	language = {en},
	number = {5471},
	urldate = {2020-02-07},
	journal = {Science},
	author = {Zaccai, Giuseppe},
	month = jun,
	year = {2000},
	pmid = {10834833},
	pages = {1604--1607},
}

@article{kim_maxima_2017,
	title = {Maxima in the thermodynamic response and correlation functions of deeply supercooled water},
	volume = {358},
	copyright = {Copyright © 2017 The Authors, some rights reserved; exclusive licensee American Association for the Advancement of Science. No claim to original U.S. Government Works. http://www.sciencemag.org/about/science-licenses-journal-article-reuseThis is an article distributed under the terms of the Science Journals Default License.},
	issn = {0036-8075, 1095-9203},
	url = {https://science.sciencemag.org/content/358/6370/1589},
	doi = {10.1126/science.aap8269},
	language = {en},
	number = {6370},
	urldate = {2020-02-04},
	journal = {Science},
	author = {Kim, Kyung Hwan and Späh, Alexander and Pathak, Harshad and Perakis, Fivos and Mariedahl, Daniel and Amann-Winkel, Katrin and Sellberg, Jonas A. and Lee, Jae Hyuk and Kim, Sangsoo and Park, Jaehyun and Nam, Ki Hyun and Katayama, Tetsuo and Nilsson, Anders},
	month = dec,
	year = {2017},
	pmid = {29269472},
	pages = {1589--1593},
}

@article{lee_microscopic_2001,
	title = {Microscopic origins of entropy, heat capacity and the glass transition in proteins},
	volume = {411},
	copyright = {2001 Macmillan Magazines Ltd.},
	issn = {1476-4687},
	url = {https://www.nature.com/articles/35078119},
	doi = {10.1038/35078119},
	language = {en},
	number = {6836},
	urldate = {2020-01-22},
	journal = {Nature},
	author = {Lee, Andrew L. and Wand, A. Joshua},
	month = may,
	year = {2001},
	pages = {501--504},
}

@article{dauchot_dynamical_2005,
	title = {Dynamical {Heterogeneity} {Close} to the {Jamming} {Transition} in a {Sheared} {Granular} {Material}},
	volume = {95},
	url = {https://link.aps.org/doi/10.1103/PhysRevLett.95.265701},
	doi = {10.1103/PhysRevLett.95.265701},
	number = {26},
	urldate = {2022-01-27},
	journal = {Phys. Rev. Lett.},
	author = {Dauchot, O. and Marty, G. and Biroli, G.},
	month = dec,
	year = {2005},
	pages = {265701},
}

@article{ruta_hard_2017,
	title = {Hard {X}-rays as pump and probe of atomic motion in oxide glasses},
	volume = {7},
	issn = {2045-2322},
	url = {http://www.nature.com/articles/s41598-017-04271-x},
	doi = {10.1038/s41598-017-04271-x},
	language = {en},
	number = {1},
	urldate = {2022-02-16},
	journal = {Sci. Rep.},
	author = {Ruta, B. and Zontone, F. and Chushkin, Y. and Baldi, G. and Pintori, G. and Monaco, G. and Rufflé, B. and Kob, W.},
	month = jun,
	year = {2017},
	pages = {3962},
}

@article{girelli_microscopic_2021,
	title = {Microscopic {Dynamics} of {Liquid}-{Liquid} {Phase} {Separation} and {Domain} {Coarsening} in a {Protein} {Solution} {Revealed} by {X}-{Ray} {Photon} {Correlation} {Spectroscopy}},
	volume = {126},
	url = {https://link.aps.org/doi/10.1103/PhysRevLett.126.138004},
	doi = {10.1103/PhysRevLett.126.138004},
	number = {13},
	urldate = {2022-02-17},
	journal = {Phys. Rev. Lett.},
	author = {Girelli, Anita and Rahmann, Hendrik and Begam, Nafisa and Ragulskaya, Anastasia and Reiser, Mario and Chandran, Sivasurender and Westermeier, Fabian and Sprung, Michael and Zhang, Fajun and Gutt, Christian and Schreiber, Frank},
	month = apr,
	year = {2021},
	pages = {138004},
}

@article{begam_kinetics_2021,
	title = {Kinetics of {Network} {Formation} and {Heterogeneous} {Dynamics} of an {Egg} {White} {Gel} {Revealed} by {Coherent} {X}-{Ray} {Scattering}},
	volume = {126},
	url = {https://link.aps.org/doi/10.1103/PhysRevLett.126.098001},
	doi = {10.1103/PhysRevLett.126.098001},
	number = {9},
	urldate = {2022-04-01},
	journal = {Phys. Rev. Lett.},
	author = {Begam, Nafisa and Ragulskaya, Anastasia and Girelli, Anita and Rahmann, Hendrik and Chandran, Sivasurender and Westermeier, Fabian and Reiser, Mario and Sprung, Michael and Zhang, Fajun and Gutt, Christian and Schreiber, Frank},
	month = mar,
	year = {2021},
	pages = {098001},
}

@article{chen_observation_2006,
	title = {Observation of fragile-to-strong dynamic crossover in protein hydration water},
	volume = {103},
	issn = {0027-8424},
	url = {https://www.ncbi.nlm.nih.gov/pmc/articles/PMC1482557/},
	doi = {10.1073/pnas.0602474103},
	number = {24},
	urldate = {2022-05-13},
	journal = {Proc. Natl. Acad. Sci. U.S.A.},
	author = {Chen, S.-H. and Liu, L. and Fratini, E. and Baglioni, P. and Faraone, A. and Mamontov, E.},
	month = jun,
	year = {2006},
	pmid = {16751274},
	pmcid = {PMC1482557},
	pages = {9012--9016},
}

@article{doster_dynamical_1989,
	title = {Dynamical transition of myoglobin revealed by inelastic neutron scattering},
	volume = {337},
	issn = {1476-4687},
	url = {https://www.nature.com/articles/337754a0},
	doi = {10.1038/337754a0},
	language = {en},
	number = {6209},
	urldate = {2022-06-09},
	journal = {Nature},
	author = {Doster, Wolfgang and Cusack, Stephen and Petry, Winfried},
	month = feb,
	year = {1989},
	pages = {754--756},
}

@article{rasmussen_crystalline_1992,
	title = {Crystalline ribonuclease {A} loses function below the dynamical transition at 220 {K}},
	volume = {357},
	copyright = {1992 Nature Publishing Group},
	issn = {1476-4687},
	url = {https://www.nature.com/articles/357423a0},
	doi = {10.1038/357423a0},
	language = {en},
	number = {6377},
	urldate = {2022-06-09},
	journal = {Nature},
	author = {Rasmussen, Bjarne F. and Stock, Ann M. and Ringe, Dagmar and Petsko, Gregory A.},
	month = jun,
	year = {1992},
	pages = {423--424},
}

@article{vitkup_solvent_2000,
	title = {Solvent mobility and the protein 'glass' transition},
	volume = {7},
	copyright = {2000 Nature America Inc.},
	issn = {1545-9985},
	url = {https://www.nature.com/articles/nsb0100_34},
	doi = {10.1038/71231},
	language = {en},
	number = {1},
	urldate = {2022-06-09},
	journal = {Nat. Struct. Biol.},
	author = {Vitkup, Dennis and Ringe, Dagmar and Petsko, Gregory A. and Karplus, Martin},
	month = jan,
	year = {2000},
	pages = {34--38},
}

@article{yang_a_2014,
	title = {A fully atomistic computer simulation study of cold denaturation of a β-hairpin},
	volume = {5},
	issn = {2041-1723},
	url = {https://www.nature.com/articles/ncomms6773},
	doi = {10.1038/ncomms6773},
	language = {en},
	number = {1},
	urldate = {2022-06-09},
	journal = {Nat. Commun.},
	author = {Yang, Changwon and Jang, Soonmin and Pak, Youngshang},
	month = dec,
	year = {2014},
	pages = {5773},
}

@article{debenedetti_second_2020,
	title = {Second critical point in two realistic models of water},
	volume = {369},
	issn = {1095-9203},
	doi = {10.1126/science.abb9796},
	language = {eng},
	number = {6501},
	journal = {Science},
	author = {Debenedetti, Pablo G. and Sciortino, Francesco and Zerze, Gül H.},
	month = jul,
	year = {2020},
	pmid = {32675369},
	pages = {289--292},
}

@article{lewandowski_direct_2015,
	title = {Direct observation of hierarchical protein dynamics},
	volume = {348},
	issn = {0036-8075},
	url = {http://www.jstor.org/stable/24747396},
	number = {6234},
	urldate = {2022-06-09},
	journal = {Science},
	author = {Lewandowski, Józef R. and Halse, Meghan E. and Blackledge, Martin and Emsley, Lyndon},
	year = {2015},
	pages = {578--581},
}

@article{ragulskaya_interplay_2021,
	title = {Interplay between {Kinetics} and {Dynamics} of {Liquid}–{Liquid} {Phase} {Separation} in a {Protein} {Solution} {Revealed} by {Coherent} {X}-ray {Spectroscopy}},
	volume = {12},
	url = {https://doi.org/10.1021/acs.jpclett.1c01940},
	doi = {10.1021/acs.jpclett.1c01940},
	number = {30},
	urldate = {2022-06-09},
	journal = {J. Phys. Chem. Lett.},
	author = {Ragulskaya, Anastasia and Begam, Nafisa and Girelli, Anita and Rahmann, Hendrik and Reiser, Mario and Westermeier, Fabian and Sprung, Michael and Zhang, Fajun and Gutt, Christian and Schreiber, Frank},
	month = aug,
	year = {2021},
	pages = {7085--7090},
}

@article{kim_experimental_2020,
	title = {Experimental observation of the liquid-liquid transition in bulk supercooled water under pressure},
	volume = {370},
	issn = {0036-8075, 1095-9203},
	url = {https://www.science.org/doi/10.1126/science.abb9385},
	doi = {10.1126/science.abb9385},
	language = {en},
	number = {6519},
	urldate = {2022-06-09},
	journal = {Science},
	author = {Kim, Kyung Hwan and Amann-Winkel, Katrin and Giovambattista, Nicolas and Späh, Alexander and Perakis, Fivos and Pathak, Harshad and Parada, Marjorie Ladd and Yang, Cheolhee and Mariedahl, Daniel and Eklund, Tobias and Lane, Thomas. J. and You, Seonju and Jeong, Sangmin and Weston, Matthew and Lee, Jae Hyuk and Eom, Intae and Kim, Minseok and Park, Jaeku and Chun, Sae Hwan and Poole, Peter H. and Nilsson, Anders},
	month = nov,
	year = {2020},
	pages = {978--982},
}

@incollection{madsen_structural_2016,
	address = {Cham},
	title = {Structural {Dynamics} of {Materials} {Probed} by {X}-{Ray} {Photon} {Correlation} {Spectroscopy}},
	isbn = {978-3-319-14394-1},
	url = {https://doi.org/10.1007/978-3-319-14394-1_29},
	language = {en},
	urldate = {2022-06-09},
	booktitle = {Synchrotron {Light} {Sources} and {Free}-{Electron} {Lasers}: {Accelerator} {Physics}, {Instrumentation} and {Science} {Applications}},
	author = {Madsen, Anders and Fluerasu, Andrei and Ruta, Beatrice},
	editor = {Jaeschke, Eberhard J. and Khan, Shaukat and Schneider, Jochen R. and Hastings, Jerome B.},
	year = {2016},
	pages = {1617--1641},
}

@incollection{grubel_x-ray_2008,
	address = {Dordrecht},
	title = {X-{Ray} {Photon} {Correlation} {Spectroscopy} ({XPCS})},
	isbn = {978-1-4020-4465-6},
	url = {https://doi.org/10.1007/978-1-4020-4465-6_18},
	language = {en},
	urldate = {2022-06-09},
	booktitle = {Soft {Matter} {Characterization}},
	author = {Grübel, G. and Madsen, A. and Robert, A.},
	editor = {Borsali, Redouane and Pecora, Robert},
	year = {2008},
	doi = {10.1007/978-1-4020-4465-6_18},
	pages = {953--995},
}

@article{sandy_hard_2018,
	title = {Hard {X}-{Ray} {Photon} {Correlation} {Spectroscopy} {Methods} for {Materials} {Studies}},
	volume = {48},
	issn = {1531-7331, 1545-4118},
	url = {https://www.annualreviews.org/doi/10.1146/annurev-matsci-070317-124334},
	doi = {10.1146/annurev-matsci-070317-124334},
	language = {en},
	number = {1},
	urldate = {2022-06-09},
	journal = {Annu. Rev. Mater. Res.},
	author = {Sandy, Alec R. and Zhang, Qingteng and Lurio, Laurence B.},
	month = jul,
	year = {2018},
	pages = {167--190},
}

@book{berne_dynamic_2000,
	title = {Dynamic {Light} {Scattering}: {With} {Applications} to {Chemistry}, {Biology}, and {Physics}},
	isbn = {978-0-486-41155-2},
	shorttitle = {Dynamic {Light} {Scattering}},
	language = {en},
	publisher = {Courier Corporation},
	author = {Berne, Bruce J. and Pecora, Robert},
	month = jan,
	year = {2000},
}

@article{williams_non-symmetrical_1970,
	title = {Non-symmetrical dielectric relaxation behaviour arising from a simple empirical decay function},
	volume = {66},
	issn = {0014-7672},
	url = {http://pubs.rsc.org/en/content/articlelanding/1970/tf/tf9706600080},
	doi = {10.1039/TF9706600080},
	language = {en},
	number = {0},
	urldate = {2022-06-10},
	journal = {Trans. Faraday Soc.},
	author = {Williams, Graham and Watts, David C.},
	month = jan,
	year = {1970},
	pages = {80--85},
}

@article{pintori_relaxation_2019,
	title = {Relaxation dynamics induced in glasses by absorption of hard x-ray photons},
	volume = {99},
	url = {https://link.aps.org/doi/10.1103/PhysRevB.99.224206},
	doi = {10.1103/PhysRevB.99.224206},
	number = {22},
	urldate = {2022-06-10},
	journal = {Phys. Rev. B},
	author = {Pintori, G. and Baldi, G. and Ruta, B. and Monaco, G.},
	month = jun,
	year = {2019},
	pages = {224206},
}

@article{madsen_beyond_2010,
	title = {Beyond simple exponential correlation functions and equilibrium dynamics in x-ray photon correlation spectroscopy},
	volume = {12},
	issn = {1367-2630},
	url = {https://doi.org/10.1088/1367-2630/12/5/055001},
	doi = {10.1088/1367-2630/12/5/055001},
	language = {en},
	number = {5},
	urldate = {2022-06-10},
	journal = {New J. Phys.},
	author = {Madsen, Anders and Leheny, Robert L. and Guo, Hongyu and Sprung, Michael and Czakkel, Orsolya},
	month = may,
	year = {2010},
	pages = {055001},
}

@article{ruta_wave-vector_2020,
	title = {Wave-{Vector} {Dependence} of the {Dynamics} in {Supercooled} {Metallic} {Liquids}},
	volume = {125},
	url = {https://link.aps.org/doi/10.1103/PhysRevLett.125.055701},
	doi = {10.1103/PhysRevLett.125.055701},
	number = {5},
	urldate = {2022-06-10},
	journal = {Phys. Rev. Lett.},
	author = {Ruta, B. and Hechler, S. and Neuber, N. and Orsi, D. and Cristofolini, L. and Gross, O. and Bochtler, B. and Frey, M. and Kuball, A. and Riegler, S. S. and Stolpe, M. and Evenson, Z. and Gutt, C. and Westermeier, F. and Busch, R. and Gallino, I.},
	month = jul,
	year = {2020},
	pages = {055701},
}

@article{schwarz_applications_1969,
	title = {Applications of the spur diffusion model to the radiation chemistry of aqueous solutions},
	volume = {73},
	issn = {0022-3654, 1541-5740},
	url = {https://pubs.acs.org/doi/10.1021/j100726a047},
	doi = {10.1021/j100726a047},
	language = {en},
	number = {6},
	urldate = {2022-06-10},
	journal = {J. Phys. Chem.},
	author = {Schwarz, Harold A.},
	month = jun,
	year = {1969},
	pages = {1928--1937},
}

@article{pawlus_conductivity_2008,
	title = {Conductivity in {Hydrated} {Proteins}: {No} {Signs} of the {Fragile}-to-{Strong} {Crossover}},
	volume = {100},
	shorttitle = {Conductivity in {Hydrated} {Proteins}},
	url = {https://link.aps.org/doi/10.1103/PhysRevLett.100.108103},
	doi = {10.1103/PhysRevLett.100.108103},
	number = {10},
	urldate = {2022-06-10},
	journal = {Phys. Rev. Lett.},
	author = {Pawlus, S. and Khodadadi, S. and Sokolov, A. P.},
	month = mar,
	year = {2008},
	pages = {108103},
}

@article{bouzid_elastically_2017,
	title = {Elastically driven intermittent microscopic dynamics in soft solids},
	volume = {8},
	issn = {2041-1723},
	url = {http://www.nature.com/articles/ncomms15846},
	doi = {10.1038/ncomms15846},
	language = {en},
	number = {1},
	urldate = {2022-06-10},
	journal = {Nat. Commun.},
	author = {Bouzid, Mehdi and Colombo, Jader and Barbosa, Lucas Vieira and Del Gado, Emanuela},
	month = aug,
	year = {2017},
	pages = {15846},
}

@article{sinha_x-ray_2014,
	title = {X-ray {Photon} {Correlation} {Spectroscopy} {Studies} of {Surfaces} and {Thin} {Films}},
	volume = {26},
	issn = {09359648},
	url = {https://onlinelibrary.wiley.com/doi/10.1002/adma.201401094},
	doi = {10.1002/adma.201401094},
	language = {en},
	number = {46},
	urldate = {2022-06-14},
	journal = {Adv. Mater.},
	author = {Sinha, Sunil K. and Jiang, Zhang and Lurio, Laurence B.},
	month = dec,
	year = {2014},
	pages = {7764--7785},
}

@article{schiro_communication:_2013,
	title = {Communication: {Protein} dynamical transition vs. liquid-liquid phase transition in protein hydration water},
	volume = {139},
	issn = {0021-9606},
	shorttitle = {Communication},
	url = {https://aip.scitation.org/doi/full/10.1063/1.4822250},
	doi = {10.1063/1.4822250},
	number = {12},
	urldate = {2022-06-30},
	journal = {J. Chem. Phys.},
	author = {Schirò, Giorgio and Fomina, Margarita and Cupane, Antonio},
	month = sep,
	year = {2013},
	pages = {121102},
}

@article{berthier_direct_2005,
	title = {Direct {Experimental} {Evidence} of a {Growing} {Length} {Scale} {Accompanying} the {Glass} {Transition}},
	volume = {310},
	issn = {0036-8075},
	url = {http://www.jstor.org/stable/3843011},
	number = {5755},
	urldate = {2022-07-01},
	journal = {Science},
	author = {Berthier, L. and Biroli, C. and Bouchaud, J. -P. and Cipelletti, L. and El Masri, D. and L'Hôte, D. and Ladieu, F. and Pierno, M.},
	year = {2005},
	pages = {1797--1800},
}

@article{mazza_more_2011,
	title = {More than one dynamic crossover in protein hydration water},
	volume = {108},
	url = {https://www.pnas.org/doi/10.1073/pnas.1104299108},
	doi = {10.1073/pnas.1104299108},
	number = {50},
	urldate = {2022-07-01},
	journal = {Proc. Natl. Acad. Sci. U.S.A.},
	author = {Mazza, Marco G. and Stokely, Kevin and Pagnotta, Sara E. and Bruni, Fabio and Stanley, H. Eugene and Franzese, Giancarlo},
	month = dec,
	year = {2011},
	pages = {19873--19878},
}

@article{mallamace_nmr_2008,
	title = {{NMR} evidence of a sharp change in a measure of local order in deeply supercooled confined water},
	volume = {105},
	url = {https://www.pnas.org/doi/full/10.1073/pnas.0805032105},
	doi = {10.1073/pnas.0805032105},
	number = {35},
	urldate = {2022-07-01},
	journal = {Proc. Natl. Acad. Sci. U.S.A.},
	author = {Mallamace, F. and Corsaro, C. and Broccio, M. and Branca, C. and González-Segredo, N. and Spooren, J. and Chen, S.-H. and Stanley, H. E.},
	month = sep,
	year = {2008},
	pages = {12725--12729},
}

@article{keys_measurement_2007,
	title = {Measurement of growing dynamical length scales and prediction of the jamming transition in a granular material},
	volume = {3},
	copyright = {2007 Nature Publishing Group},
	issn = {1745-2481},
	url = {http://www.nature.com/articles/nphys572},
	doi = {10.1038/nphys572},
	language = {en},
	number = {4},
	urldate = {2022-07-02},
	journal = {Nat. Phys.},
	author = {Keys, Aaron S. and Abate, Adam R. and Glotzer, Sharon C. and Durian, Douglas J.},
	month = apr,
	year = {2007},
	pages = {260--264},
}

@article{zhang_dynamic_2009,
	title = {Dynamic susceptibility of supercooled water and its relation to the dynamic crossover phenomenon},
	volume = {79},
	issn = {1539-3755, 1550-2376},
	url = {https://link.aps.org/doi/10.1103/PhysRevE.79.040201},
	doi = {10.1103/PhysRevE.79.040201},
	language = {en},
	number = {4},
	urldate = {2022-07-09},
	journal = {Phys. Rev. E},
	author = {Zhang, Yang and Lagi, Marco and Fratini, Emiliano and Baglioni, Piero and Mamontov, Eugene and Chen, Sow-Hsin},
	month = apr,
	year = {2009},
	pages = {040201},
}

@article{lehmkuhler_from_2021,
	title = {From {Femtoseconds} to {Hours}—{Measuring} {Dynamics} over 18 {Orders} of {Magnitude} with {Coherent} {X}-rays},
	volume = {11},
	issn = {2076-3417},
	url = {https://www.mdpi.com/2076-3417/11/13/6179},
	doi = {10.3390/app11136179},
	language = {en},
	number = {13},
	urldate = {2022-08-09},
	journal = {Appl. Sci.},
	author = {Lehmkühler, Felix and Roseker, Wojciech and Grübel, Gerhard},
	month = jan,
	year = {2021},
	pages = {6179},
}

@article{reiser_resolving_2022,
	title = {Resolving molecular diffusion and aggregation of antibody proteins with megahertz {X}-ray free-electron laser pulses},
	volume = {13},
	issn = {2041-1723},
	url = {https://www.nature.com/articles/s41467-022-33154-7},
	doi = {10.1038/s41467-022-33154-7},
	language = {en},
	number = {1},
	urldate = {2022-09-21},
	journal = {Nat. Commun.},
	author = {Reiser, Mario and Girelli, Anita and Ragulskaya, Anastasia and Das, Sudipta and Berkowicz, Sharon and Bin, Maddalena and Ladd-Parada, Marjorie and Filianina, Mariia and Poggemann, Hanna-Friederike and Begam, Nafisa and Akhundzadeh, Mohammad Sayed and Timmermann, Sonja and Randolph, Lisa and Chushkin, Yuriy and Seydel, Tilo and Boesenberg, Ulrike and Hallmann, Jörg and Möller, Johannes and Rodriguez-Fernandez, Angel and Rosca, Robert and Schaffer, Robert and Scholz, Markus and Shayduk, Roman and Zozulya, Alexey and Madsen, Anders and Schreiber, Frank and Zhang, Fajun and Perakis, Fivos and Gutt, Christian},
	month = sep,
	year = {2022},
	pages = {1--10},
}

@article{chushkin_probing_2022,
	title = {Probing {Cage} {Relaxation} in {Concentrated} {Protein} {Solutions} by {X}-{Ray} {Photon} {Correlation} {Spectroscopy}},
	volume = {129},
	url = {https://link.aps.org/doi/10.1103/PhysRevLett.129.238001},
	doi = {10.1103/PhysRevLett.129.238001},
	number = {23},
	urldate = {2022-12-12},
	journal = {Phys. Rev. Lett.},
	author = {Chushkin, Yuriy and Gulotta, Alessandro and Roosen-Runge, Felix and Pal, Antara and Stradner, Anna and Schurtenberger, Peter},
	month = nov,
	year = {2022},
	pages = {238001},
}

@article{yang_protein-water_1979,
	title = {Protein-water interactions. {Heat} capacity of the lysozyme-water system},
	volume = {18},
	issn = {0006-2960, 1520-4995},
	url = {https://pubs.acs.org/doi/abs/10.1021/bi00579a035},
	doi = {10.1021/bi00579a035},
	language = {en},
	number = {12},
	urldate = {2022-12-13},
	journal = {Biochemistry},
	author = {Yang, Pang-Hsiong and Rupley, John A.},
	month = jun,
	year = {1979},
	pages = {2654--2661},
}

@article{leung_an_1999,
	title = {An improved method for protein crystal density measurements},
	volume = {32},
	issn = {0021-8898},
	url = {http://scripts.iucr.org/cgi-bin/paper?wb0070},
	doi = {10.1107/S0021889899006809},
	language = {en},
	number = {5},
	urldate = {2022-12-13},
	journal = {J. Appl. Crystallogr.},
	author = {Leung, A. K. W. and Park, M. M. V. and Borhani, D. W.},
	month = oct,
	year = {1999},
	pages = {1006--1009},
}

@article{fujiwara_measurements_2017,
	title = {Measurements of {Thermal} {Conductivity} and {Thermal} {Diffusivity} of {Hen} {Egg}-{White} {Lysozyme} {Crystals} and {Its} {Solution} {Using} the {Transient} {Short} {Hot} {Wire} {Method}},
	volume = {38},
	issn = {1572-9567},
	url = {https://doi.org/10.1007/s10765-017-2258-y},
	doi = {10.1007/s10765-017-2258-y},
	language = {en},
	number = {8},
	urldate = {2022-12-13},
	journal = {Int. J. Thermophys.},
	author = {Fujiwara, Seiji and Maki, Syou and Maekawa, Ryunosuke and Tanaka, Seiichi and Hagiwara, Masayuki},
	month = jun,
	year = {2017},
	pages = {123},
}

@article{burtonlhenke_x-ray_1993,
	title = {X-ray interactions: photoabsorption, scattering, transmission, and reflection at {E}= 50-30,000 {eV}, {Z}= 1-92},
	volume = {54},
	number = {2},
	journal = {Atomic data and nuclear data tables},
	author = {{Burton L Henke} and {Eric M Gullikson} and {John C Davis}},
	year = {1993},
	pages = {181--342},
}

@article{albert_fifth-order_2016,
	title = {Fifth-order susceptibility unveils growth of thermodynamic amorphous order in glass-formers},
	volume = {352},
	url = {https://www.science.org/doi/10.1126/science.aaf3182},
	doi = {10.1126/science.aaf3182},
	number = {6291},
	urldate = {2022-12-19},
	journal = {Science},
	author = {Albert, S. and Bauer, Th. and Michl, M. and Biroli, G. and Bouchaud, J.-P. and Loidl, A. and Lunkenheimer, P. and Tourbot, R. and Wiertel-Gasquet, C. and Ladieu, F.},
	month = jun,
	year = {2016},
	pages = {1308--1311},
}

@article{roh_dynamics_2009,
	title = {Dynamics of {tRNA} at {Different} {Levels} of {Hydration}},
	volume = {96},
	issn = {0006-3495},
	url = {https://www.sciencedirect.com/science/article/pii/S0006349509003129},
	doi = {10.1016/j.bpj.2008.12.3895},
	language = {en},
	number = {7},
	urldate = {2022-12-20},
	journal = {Biophys. Chem.},
	author = {Roh, J. H. and Briber, R. M. and Damjanovic, A. and Thirumalai, D. and Woodson, S. A. and Sokolov, A. P.},
	month = apr,
	year = {2009},
	pages = {2755--2762},
}

\end{document}